\documentclass[12pt]{article}
\usepackage{latexsym}
\usepackage{amssymb}
\usepackage{amsfonts}
\usepackage{pslatex}
\usepackage{graphicx}

\catcode`\@=11
\def\numberbysection{\@addtoreset{equation}{section}
        \def\theequation{\thesection.\arabic{equation}}}

\def\be{\begin{equation}}
\def\ee{\end{equation}}
\def\ba{\begin{eqnarray}}
\def\ea{\end{eqnarray}}

\def\nl{\nonumber \\}

\def\Tr{{\rm Tr}}
\def\dag{\dagger}

\def\D{\Delta}

\def\eps{\varepsilon}

\def\th{\theta}

\def\w{\omega}

\numberbysection

\begin{document}

\begin{titlepage}
\begin{center}


\vspace{1cm}

{\Large \bf Edge excitations of the  } \\
\vspace{.4cm}
{\Large \bf Chern Simons matrix theory for the FQHE. }\\

\vspace{1cm}

Ivan D. RODRIGUEZ\footnote{Electronic address: rodriguez@fi.infn.it}\\
\medskip
{\em I.N.F.N. and Dipartimento di Fisica} \\
{\em  Via G. Sansone 1, 50019 Sesto Fiorentino - Firenze, Italy} \\\end{center}
\vspace{.5cm}
\begin{abstract}
We study the edge excitations of the Chern Simons matrix theory, describing the Laughlin fluids for filling fraction $\nu=\frac{1}{k}$, with $k$ an integer. Based on the semiclassical solutions of the theory, we are able to identify the bulk and edge degrees of freedom. In this way we can freeze the bulk of the theory, to the semiclassical values, obtaining an effective theory governing the boundary excitations of the Chern Simons matrix theory. Finally, we show that this effective theory is equal to the chiral boson theory on the circle.

\end{abstract}

\vfill
\end{titlepage}
\pagenumbering{arabic}

\section{Introduction}

In this paper we study the edge excitations of the regularized
non-commutative Chern Simons(CS) matrix theory \cite{susskind}\cite{poly1}.
In Ref.\cite{susskind} Susskind showed that the theory of an incompressible fluid of charged particles in an strong
magnetic field $B$ corresponds, in the semiclassical regime, to the small $\theta$ limit of the non-commutative CS theory \cite{noncomm}, where $\theta$ is the non-commutative parameter.  Moreover, the classical theory of the fluid presents a symmetry of diffeomorphisms that preserve the area and thus
it can be mapped to the non-commutative CS matrix theory, by means of the Goldstone-Hoppe regularization\cite{hoppe}. Another approach that relates the fluid theory with the non-commutative CS theory comes from
the analogies between the physics of electrons in an strong magnetic field and
the properties of D-branes in String Theory \cite{susskind}\cite{dzero}. Based on the previous arguments, Susskind proposed the non-commutative CS theory to describe the Laughlin Hall states \cite{laugh}\cite{prange} made of particles with an effective area \footnote{Note that in a non-commutative space parameterized by coordinates $y^1$ and $y^2$ we have $\left[y^1,y^2 \right]=\theta$, where $\theta$ can be thought as a unity of area.}.

The non-commutative CS theory is an $U(N)$ gauge theory in one dimension (i.e. time)
with CS kinetic term; it involves two hermitian matrix coordinates $X_1(t)$ and $X_2(t)$,
that are non-commuting, $\left[ X_1,X_2 \right] = i \theta I$, where $I$ is the identity
and the quantity $B\theta=k$ is quantized to an integer, at quantum level, to preserve gauge invariance \cite{susskind}. As we will show in section 2.1, the eigenvalues of the matrices describe the coordinates of the particles\cite{susskind}\cite{poly1}. Also the parameter $k$ is related to the filling fraction of the fluid of electrons as $\nu=\frac{1}{k}$\cite{susskind}.

Although Susskind theory shares a similar behavior with the fractional quantum Hall effect (FQHE), it contains an infinite number of degrees of freedom (d.o.f.). Due to this, Polychronakos regularized the theory to allow for a finite number of particles \cite{poly1}. The quantum ground state of the Susskind-Polychronakos theory was found to be the Laughlin wave function for the FQHE, with filling fraction $\nu=\frac{1}{k}$ \cite{cr}-\cite{frad}.

As said before, in this work we will study the edge excitations of the Susskind-Polychronakos matrix theory.
We remark that the boundary excitations of the FQHE in the context of non-commutative theories, have been studied in the past from other points of view \cite{cabra}\cite{berens}. The paper is organized as follows: in section 2 we briefly review the non-commutative CS theory. We introduce the Susskind theory and the regularization by Polychronakos to allow for matrices of finite order. In section 2.1, we show that the
quantum ground state of the regularized matrix theory corresponds to the Laughlin wave function with filling $\nu=\frac{1}{k}$ \cite{cr}-\cite{frad}.
In the last part of the section we show that the low lying excitations of the ground state (edge excitations) are in one to one correspondence with the excitations of the chiral boson theory, describing the boundary physics of the Laughlin fluids \cite{wen}\cite{winf}\cite{cftheories}. In section 2.2 we present the semiclassical ground state solution obtained by Polychronakos in \cite{poly1}. In section 3.1 we take into account the more general excitations of the semiclassical ground state solution, that produce only a perturbation on the edge of the droplet, i.e. that leave the bulk unchanged. This means that these perturbations don't introduce any interactions between the bulk and the boundary. In this way we are able to obtain the relevant d.o.f. on the edge. Based on the previous result, we introduce a mean field approximation that freezes the bulk d.o.f. and their interactions with the boundary of the droplet. Therefore we obtain an effective theory formulated only in terms of the dynamics of the edge d.o.f. In section 3.2, we show that this effective theory corresponds to the chiral boson theory on the circle that, as said before, describes the edge excitations of the Laughlin fluids \cite{wen}\cite{winf}. This is the main result of the paper. Finally we make some comments about the application of our analysis to the Maxwell CS matrix theory \cite{we}\cite{we2} for the Jain hierarchy \cite{jainbook}.

\section{Review of the non-commutative Chern Simons theory}

In an interesting paper \cite{susskind} Susskind conjectured that the non-commutative Chern
Simons field theory in two dimensions could describe the Laughlin incompressible fluids
in the QHE. This conjecture was inspired by the fact that the semiclassical
limit of this theory \cite{sussk2} describes incompressible fluids in high magnetic field with Laughlin's
filling fractions ($\nu = \frac{1}{k}$, $k$ an integer) and their quasi-hole excitations.

The non-commutative CS theory can be represented as a matrix theory given by \cite{susskind}:
\ba
S_{Susskind} = \int \ dt\ \frac{B}{2}
\Tr\left[ \ \eps_{ij} \ X_i(t)\ D_t \ X_j(t) +\ 2\th\ A_0(t) \right] , \label{mcs-action1}
\ea
where $X_1(t)$ , $X_2(t)$ and $A_0(t)$ are $N \times N$ hermitian matrices, with $N=\infty$, and the covariant derivative is defined as $D_t X_j=\dot{X}_j -i \left[ A_0,X_j \right]$. The parameter $\theta$ is related to the density of the fluid of electrons $\rho_0$ as:
\ba
\rho_0=\frac{1}{2 \pi \theta} .
\label{dens}
\ea
In this theory the Gauss law ($\frac{\partial L}{\partial A_0}=0$) implies:
\ba
\left[ X_1,X_2\right] = i \theta I \ ,
\label{GaussSuss}
\ea
where $I$ is the identity matrix. Taking the trace in both members of (\ref{GaussSuss}) it is evident that it is satisfied only for $N=\infty$. Therefore Susskind's theory applies to an infinite system. Instead the FQHE is a system with a boundary and a finite number of particles. For this reason Polychronakos \cite{poly1} proposed the following action generalizing the Susskind theory:
\ba
S=\int dt \frac{B}{2} Tr\left \{
\epsilon_{ab} (\dot{X}_a + i [A_0,X_a] ) X_b + 2 \theta A_0 - \omega X_a^2 \right \} \ + \int dt \psi^\dag(i\dot{\psi}-A_0
\psi) . \ \nl \label{CS-poly}
\ea
He adds two new terms to Susskind's action (\ref{mcs-action1}). The first term is an harmonic oscillator potential for
the matrices that confines the eigenvalues, i.e. that keeps the particles localized in the plane. The second term is proportional to a complex N-vector $\psi$.

The Gauss law is now given by:
\ba
G \equiv \ -i \ B [ X_1 , X_2] + \psi \psi^\dag - B \theta I = 0 . \ \label{poly-glaw}
\ea
Observe that the trace of (\ref{poly-glaw}) implies,
\ba
\psi^\dag \psi = N B \theta, \ \label{trace-glaw}
\ea
that can be realized with finite dimensional matrices. In this way it is possible to localize, by means of the potential, a finite number of d.o.f. in the plane and therefore to introduce a boundary in the theory.

The Chern Simons theory (\ref{CS-poly}) has the $U(N)$ symmetry:
\ba
&& X_a \rightarrow U X_a U^\dag \ , \qquad \psi
\rightarrow U \psi , \nl && A_0 \rightarrow U A_0 U^\dag - i U \frac{d U^\dag}{dt} \ . \ \label{gaugetransf}
\ea
Under a gauge transformation (\ref{gaugetransf}), the action (\ref{CS-poly}) changes by the winding number of the group
element U,
\ba
S \rightarrow S - i B \theta \int dt Tr \left[ U^\dag \dot{U}  \right] ,
\ea
and gauge invariance is satisfied, at quantum level, if $ B \theta=k $ is an integer \cite{nair}.

Note that the equation of motion for $\psi$ in the $A_0=0$ gauge implies $\dot{\psi}=0$: it is an auxiliary field with trivial dynamics.

\subsection{Covariant quantization}

In the following we will work in holomorphic coordinates $X=X_1+iX_2$ and $\bar{X}=X_1 - i X_2$, with the bar denoting the Hermitian conjugate of classical matrices.

Before quantizing the Chern Simons matrix theory, we express (\ref{CS-poly}) in terms
of holomorphic matrices, in the $A_0=0$ gauge:
\ba
S &=& \int dt \left( \frac{B}{2i} \sum_{n,m=0}^N \dot{X}_{nm}
\bar{X}_{mn} + i \sum_{n=0}^N \dot{\psi}_n \psi^\dag_n - \frac{B\omega}{2} \sum_{n,m=0}^N \bar{X}_{nm} X_{mn} \right) \ , \nl G &=& -
\frac{B}{2} [ \bar{X} , X ] + \psi \psi^\dag - B \theta = 0 . \ \label{CS-poly-holom}
\ea
The form of the action (\ref{CS-poly-holom}) is that of $( (N+1)^2 + N+1 )$ particles in the lowest Landau level with
coordinates $X_{nm}$ and  $\psi_n$.

The canonical commutation relations are given by \cite{cr}:
\ba
\left[ \left[ \bar{X}_{ij}, X_{kl} \right] \right] &=&
\frac{2}{B} \delta_{jk} \delta_{il} \ , \nl \left[ \left[ \bar{\psi}_i,\psi_j \right] \right] &=&
\delta_{ij} \ . \ \label{commutation-holom}
\ea
The double brackets are used to denote the quantum mechanical commutators between matrix elements. We use the standard polarization in quantum mechanics, i.e. the canonical conjugate momentum becomes:
\ba
\bar{X}_{nm} \rightarrow \frac{2}{B}
\frac{\partial}{\partial X_{mn}}, \qquad \bar{\psi}_n\rightarrow \frac{\partial}{\partial \psi_n} \ , \
\label{polarization}
\ea
and using (\ref{polarization}), it can be shown that the Gauss law implies that the physical states must be singlet of $U(N)$ made of matrices $X$ and having a number of vector $\psi$'s equal to $(N+1)k$ \cite{cr}\cite{heller}.

The general solution of the Gauss law constraint has been found in Ref. \cite{heller}. A complete basis is given by:
\ba
\Phi(X,\phi)&=&\Phi_{ \{ n_1^1,...,n_N^1 \} }...\Phi_{ \{
n_1^k,...,n_N^k \} } \qquad \mbox{with} \nl \Phi_{ \{ n_1^j,...,n_N^j \} } &=&
\epsilon^{i_1...i_N}(X^{n_1^j}\psi)_{i_1}...(X^{n_N^j}\psi)_{i_N}, \quad 0 \leq n_1^j < n_2^j < ... < n_N^j \ .
\label{physical-states}
\ea
To find the ground state of the theory observe that the Hamiltonian in (\ref{CS-poly-holom}), $\frac{B\omega}{2} Tr (\bar{X} X)$, basically counts the number of $X$ matrices (due to the harmonic oscillator commutators (\ref{commutation-holom})) appearing in the wave functions (\ref{physical-states}). Therefore the ground state corresponds to the state with the lowest number of $X$ matrices. It is given by \cite{cr}:
\ba
\Phi_{k-gs} = \left [
\epsilon^{i_1...i_N}\psi_{i_1}(X\psi)_{i_2}...(X^{N-1}\psi)_{i_N} (X^{N}\psi)_{i_{N+1}} \right ]^k . \
\label{laughlin-wf}
\ea
Note that any other state with lower number of $X$ matrices will be zero, due to the antisymmetry of the $\epsilon$ tensor.

In Ref.\cite{heller} is presented an equivalent basis, in which the states are factorized into the ground state (\ref{laughlin-wf}) and the "bosonic" powers of $X$, $\Tr \left( X^{m_i}\right)$,
with positive integers $\{m_1,\dots,m_k\}$ unrestricted, i.e.
\ba
\Phi \left(X,\psi\right) &=& \sum_{\{m_k\}}\ \Tr \left(X^{m_1}\right)\cdots
\Tr \left(X^{m_k}\right) \ \Phi_{k-gs}\ .
\label{HVR-bose}
\ea
Let us now perform the change of variables given by \cite{cr}:
\ba
X &=& V^{-1} \Lambda V, \qquad
\Lambda=diag(\lambda_0,....,\lambda_N), \nl \psi &=& V^{-1} \phi  \ , \
\label{gauge-diagonal}
\ea
on the ground state (\ref{laughlin-wf}):
\ba
\Phi_{k-gs}(\Lambda,V,\psi)&=&\left[
\epsilon^{i_0...i_N}(V^{-1}\phi)_{i_0}(V^{-1}\Lambda\phi)_{i_1}...(V^{-1}\Lambda^{N}\phi)_{i_{N+1}} \right ]^k \nl
&=&\left[ (det V)^{-1} det(\lambda_j^{i-1}\phi_j) \right]^k \nl &=& (det V)^{-k} \prod_{0 \leq n \leq m \leq N}
(\lambda_n-\lambda_m)^k \left( \prod_i \phi_i \right)^k . \
\label{laughlin-wf-eigen}
\ea
We obtain the Laughlin wave function as ground state of the Chern Simons theory, with the coordinates of the electrons
identified with the eigenvalues of the $X$ matrix\footnote{In Laughlin theory, the exponent $k$ in (\ref{laughlin-wf-eigen}) is related to the filling fraction as $\nu=\frac{1}{k}$. Nevertheless in the matrix theory, due to (\ref{gauge-diagonal}), appears a Vandermonde in the measure of integration that corrects the filling fraction in 1, i.e. $\nu=\frac{1}{k+1}$ \cite{cr}.}. It can be shown that the dependence on $\phi$ and $V$ in (\ref{laughlin-wf-eigen}) is the same for all
the physical states and so we can drop their contribution \cite{cr}. Equation (\ref{laughlin-wf-eigen}) is the most important result of the Susskind-Polychronakos theory.

The excitations of (\ref{laughlin-wf}) correspond to quasi-hole solutions in the matrix theory \cite{susskind}\cite{poly1}\cite{cr}\cite{heller}, i.e. gapful localized  density deformations (see eq. (\ref{quasip-inorig}) in the next section). Multiplying the wave function by polynomials of $\Tr(X^r)$ as in (\ref{HVR-bose}), we find  states with energy given by the boundary potential, $\D E=\w r$. These
are the basis of holomorphic excitations over the Laughlin state.
For $r=O(1)$ ($\omega\sim\frac{B}{N}$) we obtain the low energy excitations of the CS matrix theory with energy  $\D E= O(r\ B/N)$. They correspond to quasi-holes in the origin of the fluid and are in one to one correspondence with the excitations of the chiral boson theory \cite{cr}\cite{sakita}\cite{wen}\cite{winf}.

On the other hand, the quasi-particle excitation cannot be
realized in the Chern Simons matrix theory \cite{poly1}\cite{we}\cite{we2}.

\subsection{Semiclassical solutions}

In this section we study the classical solutions of the CS matrix theory (\ref{CS-poly-holom}) for $N>>1$, i.e. the semiclassical regime \cite{poly1}\cite{hans1}\cite{we2}. These solutions were found by Polychronakos in Ref.\cite{poly1}.

The Hamiltonian of the theory is given by:
\ba
H=\frac{\omega B}{2} Tr X \bar{X} + Tr \
\Lambda (- \frac{B}{2} [ \bar{X} , X ] + \psi \psi^\dag - B \theta ) \ , \
\label{poly-hamilt}
\ea
where we introduced the Gauss law constraint by means of the Lagrange multiplier $\Lambda$.

The equations of motion are given by:
\ba
\dot{X}_{ba}&=&\frac{\partial H}{\partial \Pi_{ba}}=\omega X_{ba} +
\frac{2}{B} [ \Lambda , X ]_{ba} , \nl \dot{\Pi}_{ba}&=& -\frac{\partial H}{\partial X_{ab}} =
\frac{B\omega}{2} \bar{X}_{ba} + [ \bar{X}, \Lambda ]_{ba} \ ,
\label{poly-eqmotion2}
\ea
with the canonical conjugate momentum $\Pi_{ab}=\frac{B}{2} \bar{X}_{ba}$. The minimum of energy must
satisfies the Gauss law and the equations of motion,
\ba
&& G \equiv \ -  \frac{B}{2} [ \bar{X} , X] + \psi \psi^\dag - B \theta = 0 , \nl && [
\Lambda,\bar{X} ]_{ba} = \frac{\omega B}{2} \bar{X}_{ba}.
\label{eq-min}
\ea
These are the commutation relations for a (truncated) quantum harmonic oscillator, with $\Lambda$ playing the role of
the Hamiltonian. The solutions of (\ref{eq-min}) (setting $B=2$) in the gauge in which $ \psi=\sqrt{2\theta (N+1)} \ \mid N > $ are \cite{poly1}:
\ba
\bar{X} \ = \ \sqrt{2 \theta} \
\sum_{n=0}^{N} \sqrt{n} \mid n > < n - 1 \mid \ \ , \quad \ \Lambda  = \omega \ \sum_{n=0}^{N} n \mid n > <
n \mid \ \  \ .
\label{solution-poly}
\ea
In the large $N$ limit, solution (\ref{solution-poly}) corresponds to a circular quantum Hall droplet of radius $\sqrt{2N \theta}$
\cite{poly1}. The radius-squared matrix coordinate $R^2$ is diagonal, and given by:
\ba
R^2 \ = \frac{1}{2} \left( \bar{X} X + X \bar{X} \right) = {\rm diag}\left(\theta,3\theta,5\theta,7\theta,\dots,(2N-1)\theta, N\theta \right)\ .
\label{polyr1}
\ea
From the distribution of the eigenvalues in (\ref{polyr1}) it is clear that in the $N >>1$ limit, this solution implies a droplet of constant density. Also in this limit, the filling fraction is the Laughlin value; according to the identification of the $\theta$ parameter (\ref{dens}),
\ba
\nu =\frac{2 \pi \rho_0}{e B}=\frac{1}{k} \ , \quad \rho_0= \frac{1}{2\pi\theta} \ . \
\label{Laugh-fill1}
\ea
As said before, Polychronakos theory does not contain quasi-particle excitations, only quasi-holes are present \cite{poly1}. For example, a quasi-hole of charge $-q$ in the origin of the droplet is given by (setting $B=2$),
\ba
\bar{X} = \sqrt{2 \theta} \left( \sqrt{q} \mid 0 \rangle \langle N
\mid + \sum_{n=1}^{N} \sqrt{n+q} \mid n \rangle \langle n - 1 \mid \right), \quad q>0.
\label{quasip-inorig}
\ea
It is easy to check that the $R^2$ matrix corresponding to solution (\ref{quasip-inorig}) is diagonal with eigenvalues:
\ba
R^{2(q)} \ = {\rm diag}\left((1+2q)\theta,(3+2q)\theta,(5+2q)\theta,(7+2q)\theta,\dots,(2N-1+2q)\theta, (N+2q) \theta \right)\ , \nl
\label{polyr2}
\ea
where $(q)$ indicates the charge of the quasi-hole.

\section{Effective theory on the boundary}

\subsection{Mean field approximation}

In this section we identify the relevant d.o.f. on the edge of the droplet (\ref{polyr1}) and we perform a mean field approximation that freezes the bulk d.o.f. in the CS theory (\ref{CS-poly-holom}), obtaining an effective theory for the boundary.

The starting point is the CS matrix theory (\ref{CS-poly-holom}) with Lagrangian:
\ba L_{CS}= \frac{B}{2i}
\sum_{n,m=0}^N \dot{X}_{nm} \bar{X}_{mn} + i \sum_{n=0}^N \dot{\psi}_n
\psi_n^\dag - \frac{B}{2} \omega \left( \sum_{n,m=0}^N \bar{X}_{nm} X_{mn} \right)^2 \ ,
\label{LCS}
\ea
and Gauss law constraint:
\ba
G= - \frac{B}{2} \left[ \bar{X},X \right] +
\psi \psi^\dag - B \theta = 0
\label{GaussCS}  \ .
\ea
Note that in (\ref{LCS}) we are considering the square of the potential
(\ref{CS-poly}, \ref{CS-poly-holom}), introduced by Polychronakos. We consider this potential because, as will be clear later, it gives dynamics to the edge excitations. It is easy to prove that the quantum and semiclassical results of the previous sections remain unchanged by this modification. To see this, from the quantum point of view, observe that the Polychronakos Hamiltonian, $H= \frac{B \omega}{2}Tr \left( \bar{X}X \right)$, is basically a number operator (due to the harmonic oscillator commutators (Eq.\ref{commutation-holom})) counting the number of $X$ matrices $N_X$ appearing in the wave functions (\ref{physical-states}). Therefore the Hamiltonian (\ref{LCS}):
\ba
\frac{B \omega}{2} \left( Tr \left( X \bar{X} \right) \right)^2 \ ,
\label{newpot}
\ea
corresponding to the square of the Polychronakos potential, admits the same eigenfunctions of  Polychronakos theory (given by the complete basis (\ref{physical-states}, \ref{HVR-bose})), but with energy proportional to $N_X^2$. At last, to prove that the semiclassical solutions of the CS matrix theory (section 2.2) remain unchanged with the potential (\ref{newpot}), we must to compare the equations of motion of our theory:
\ba
\dot{X}_{ba}= \omega  Tr\left( \bar{X}X \right) X_{ba} \ + \frac{2}{B} \left[ \Lambda, X \right]_{ba} \quad ; \quad \mbox{c.c.} \ ,
\label{eqMotion}
\ea
with those of Polychronakos (\ref{poly-eqmotion2}). Observe that the only difference is given by the conserved quantity $Tr\left( \bar{X}X \right)$ appearing in the second member of (\ref{eqMotion}). If we now define a new frequency, $\omega'=\omega \Tr\left( \bar{X}X \right)$, we recover the equations of motion of the CS matrix theory (\ref{poly-eqmotion2}).

From the Lagrangian (\ref{LCS}) it is clear that the field $\psi$ is a non-dynamical variable (see section 2), i.e. $\dot{\psi}=0$. Therefore in the following by means of a time-independent gauge transformation we will fix $\psi$ to the semiclassical value (see section 2.2):  $\psi=( 0,..,\sqrt{2(N+1)\theta})$.
Clearly this gauge fixing has a trivial Fadeev-Popoov
term. Thus, the only change in the path integral is in the Gauss law constraint that becomes:
\ba
G_{\alpha\beta}=0 \Rightarrow \left[
X, \bar{X}\right]_{\alpha\beta} = 2 \theta \delta_{\alpha\beta} - 2(
N +1)\theta \delta_{\alpha N}\delta_{\beta N} \ , \ \alpha , \beta
=0,...,N \ .
\label{Gaussfix}
\ea
In (\ref{Gaussfix}) and in the rest of the paper we will set $B=2$.

Now we will start with the study of the edge excitation of the theory (\ref{LCS}, \ref{GaussCS}). First of all we need to identify the relevant d.o.f. on the boundary of the droplet. To do this we can consider a perturbation of the semiclassical ground state solution (\ref{solution-poly}) that leaves all eigenvalues of $R^2$ invariant, except the last one. The only possibility is:
\ba
R^2_{\alpha\beta} = \frac{1}{2} \left( \langle\bar{X}X \rangle_{\alpha \beta}+ \langle X \bar{X} \rangle_{\alpha \beta} \right) \left( 1- \delta_{\alpha N}\delta_{\beta N} \right) +  \frac{1}{2}\left( \left( \bar{X}X \right)_{\alpha\beta} + \left( X \bar{X} \right)_{\alpha \beta} \right) \delta_{\alpha N} \delta_{\beta N} ,
\label{meanfield1}
\ea
or more clearly:
\ba
R^2 = Diag \left( \theta,3\theta,5\theta,...,(2N-1)\theta,\frac{1}{2} \left\{\bar{X},X \right\}_{NN} \right) .
\label{meanfield}
\ea
In (\ref{meanfield1}) the brackets $\langle O \rangle$ indicate the quantity $O$ valued in the semiclassical ground state solution (\ref{solution-poly}). Note that any perturbation on the off-diagonal elements of $R^2$ that affects the last eigenvalue, also produces a change in the bulk eigenvalues. Such a perturbation can be considered as a kind of interaction between the bulk and the boundary, but we are not interested in them.

Therefore (\ref{meanfield1}, \ref{meanfield}) gives us the relevant d.o.f. on the boundary and so we will consider it as our mean field approximation, i.e. we will freeze the bulk d.o.f. to the semiclassical value:
\ba
\left(\bar{X}X\right)_{\alpha \beta} = \langle \bar{X} X \rangle_{\alpha \beta} \ , \ \left(X\bar{X}\right)_{\alpha \beta} = \langle X \bar{X} \rangle_{\alpha \beta} \ ; \ \forall \ \alpha , \beta \ / \ \alpha \neq N \wedge \beta \neq N \ .
\label{meanfield2}
\ea
Now, introducing (\ref{meanfield2}) in the Gauss law constraint (\ref{Gaussfix}), we obtain:
\ba
G_{\alpha\beta} &=& \left( \langle X \bar{X}  \rangle_{\alpha\beta} - \langle \bar{X} X \rangle_{\alpha\beta} - 2\theta \delta_{\alpha \beta} \right)\left(1-\delta_{\alpha N}\delta_{\beta N} \right) + \nl && \left( \left[ X,\bar{X} \right]_{\alpha\beta} + 2N\theta \delta_{\alpha\beta} \right)\delta_{\alpha N}\delta_{\beta N} = 0 \ ; \ \alpha, \beta=0,...,N \ .
\label{Gauss4}
\ea
Note that the first term in (\ref{Gauss4}) is equal to zero because the matrices $\langle \bar{X} X \rangle$ and $\langle X \bar{X}  \rangle$ are valued in the semiclassical ground state solution (\ref{solution-poly}) that verifies (\ref{Gaussfix}). Finally we obtain the following constraint for the boundary d.o.f.:
\ba
G_{NN}=\sum_{\alpha=0}^N \left( X_{N \alpha} \overline{X}_{\alpha N}
- \overline{X}_{N \alpha} X_{\alpha N} \right)  + 2 \theta N \ .
\label{Gaussedge}
\ea

At last, if we introduce the mean field (\ref{meanfield2}) into the CS Lagrangian (\ref{LCS}) we obtain the following effective Lagrangian on the boundary:
\ba
L_{CS}&=&-i \sum_{\alpha} \left( \bar{X} \dot{X} \right)_{\alpha \alpha} - \frac{\omega}{2} J_0 \left( \left( \bar{X} X \right)_{NN} + \left( X \bar{X} \right)_{NN} \right)
 -\nl && \frac{\omega}{4} \left( \left( \bar{X}X \right)_{NN} + \left( X \bar{X} \right)_{NN} \right) ^ 2 -  \frac{\omega}{4} J_0^2 \ ,
\label{LCSmeanfield}
\ea
with $J_0=\sum_{i=0}^{N-1} \left( \langle  \bar{X}X \rangle_{ii} + \langle  X\bar{X} \rangle_{ii} \right)$. From (\ref{LCSmeanfield}) it is clear that the only variables that propagate are those corresponding to the matrix elements $\bar{X}_{\alpha N}$,  $\bar{X}_{N \alpha}$, $X_{\alpha N }$ and  $X_{N \alpha}$ and thus we can integrate out all the other coordinates in the path integral. Finally we obtain the following effective theory, $Z_b$, on the edge:
\ba
Z_b=\int DX_{N\alpha} DX_{\alpha N} D\overline{X}_{N
\alpha } D\overline{X}_{\alpha N} e^{-S_b} \delta \left( G_{NN} \right) \ ,
\label{edgeEff}
\ea
with
\ba
S_b&=& \int dt L_b = \int dt \left( \frac{1}{i} \sum_{\alpha=0}^{N-1}
\left( \dot{X}_{N \alpha} \overline{X}_{\alpha N} + \dot{X}_{\alpha
N} \overline{X}_{N \alpha} \right) -\frac{\omega}{2} J_0  \sum_{\alpha=0}^{N-1} \left(
X_{N\alpha} \overline{X}_{\alpha N} + X_{\alpha N} \overline{X}_{N
\alpha} \right)  - \right. \nl  && \left. \frac{\omega}{4} \left( \sum_{\alpha=0}^{N-1}  \left(
X_{N\alpha} \overline{X}_{\alpha N} + X_{\alpha N} \overline{X}_{N
\alpha} \right) \right) ^2 \right) \ ,
\label{otraS}
\ea
where $G_{NN}$ in (\ref{edgeEff}) is given by (\ref{Gaussedge}).
Note that in (\ref{otraS}) we have not considered the contribution of the elements $X_{NN}$ and $\bar{X}_{NN}$. They are not present in the constraint (\ref{Gaussedge}) and correspond to an ambiguity in the definition of the discrete fields $X_{\alpha N}$, $X_{N \alpha}$ and their corresponding hermitian conjugate.

\subsection{Effective edge theory equal to chiral boson theory}

In this section we show that the effective theory (\ref{edgeEff}) describing the boundary physics of the CS matrix theory corresponds to the chiral boson theory on the circle.

In the following we will work in real coordinates, considering the change of variables:
\ba
X_{N\alpha}=\sqrt{2 \theta} P^{1/2}_{\alpha} e^{i R_{\alpha}} \ , \overline{X}_{\alpha
N}=\sqrt{2 \theta} P^{1/2}_{\alpha}e^{-i R_{\alpha}} \ ,   X_{\alpha N}=\sqrt{2 \theta} Q^{1/2}_{\alpha}e^{i T_{\alpha}} \ , \overline{X}_{N \alpha
}=\sqrt{2 \theta} Q^{1/2}_{\alpha}e^{-i T_{\alpha}}
\ . \nl
\label{newvar}
\ea
Later it will be clear why we introduce the factor $\sqrt{2\theta}$ in (\ref{newvar}). In terms of these new coordinates the Lagrangian (\ref{otraS}) and the constraint (\ref{Gaussedge}) are given by:
\ba
L_b = 2\theta \sum_{\alpha=0}^{N-1} (P_\alpha \dot{R}_\alpha + Q_\alpha \dot{T}_\alpha) - J_0 \omega\theta \sum_{\alpha=0}^{N-1} (P_\alpha + Q_\alpha) - \theta^2 \omega \left( \sum_{\alpha=0}^{N-1} (P_\alpha + Q_\alpha) \right)^2 \
\label{GaussBound}
\ea
and
\ba
G_{NN} =  \sum_{\alpha=0}^{N-1}
\left( P_\alpha - Q_\alpha  \right) +  N  = 0 \ .
\label{GaussBound1}
\ea
We can interpret the coordinates $P_\alpha,Q_\alpha,R_\alpha$ and $T_\alpha$ as the d.o.f. of fields $P,Q,R$ and $T$ defined on a lattice $\alpha=0,1,..,N-1$. Because these fields describe the physics on
the edge of the droplet, i.e. on a circle of radius $R \simeq \sqrt{2N\theta}$ (see section 2.2),  it is natural to assume a periodic lattice with the points $0$ and $N-1$ identified. Below we shall see that with this assumption we get a natural
description of the theory in terms of a chiral boson defined on a circle and
with the expected relation between charge and winding number.

In the previous section, to freeze the bulk d.o.f. in the CS theory, we have implemented the mean field approximation (\ref{meanfield2}) in the particular case in which non quasi-holes are present in the bulk and therefore our study is limited only to neutral boundary excitations of the droplet. However the FQHE presents also charged edge excitations. They can be obtained creating quasi-holes in the origin of the fluid which correspond to the low energy excitations of the CS matrix theory \cite{poly1}\cite{we2}. Therefore, to consider charged boundary excitations in our effective theory (\ref{GaussBound}, \ref{GaussBound1}) we must to evaluate the mean field approximation (\ref{meanfield2}) not in the ground state solution (\ref{solution-poly}) but in the semiclassical solution (\ref{quasip-inorig}) of a quasi-hole in the origin of the fluid. On the edge d.o.f. we will consider that the effect of a quasi-hole in the bulk will be a nontrivial winding number on the fields $P,Q,R$ and $T$, i.e.:
\ba
P_{N-1}=P_0 + \alpha_1 \ , \ Q_{N-1}=Q_0 + \alpha_2 \ , \ R_{N-1}=R_0 + \beta_1 \ , \ T_{N-1}=T_0 + \beta_2 \ ,
\label{bound}
\ea
where $\alpha_1$, $\alpha_2$, $\beta_1$ and $\beta_2$ are constants. In conclusion, in the presence of bulk quasi-holes in the CS matrix theory (\ref{CS-poly-holom}) the only change in our effective theory (\ref{GaussBound}, \ref{GaussBound1}) will be in the constant $J_0$, in (\ref{GaussBound}), that now will be valued in the quasi-hole solution (\ref{quasip-inorig}) and in that the fields $P$, $Q$, $R$ and $T$ will satisfy the boundary conditions (\ref{bound}).

At this point it is interesting to see how are related the winding numbers (\ref{bound}), in the boundary theory, with a quasi-hole in the bulk theory. First of all note that in presence of a quasi-hole of charge $q$ in the origin of the droplet (\ref{quasip-inorig}), the eigenvalues of the $R^2$ matrix (\ref{polyr2}) satisfy  $R^{2 (q)}_{\alpha}=R^{2 (0)}_{\alpha} + 2 \theta q$ (the quantity in the parentheses indicates the charge of the quasi-hole) and in particular:
\ba
R^{2  (q)}_{N} =   R^{2  (0)}_{N} + 2 \theta q \ .
\label{rel2}
\ea
Observe also that using the constraint (\ref{GaussBound1}) we can write $R^2_{N}$ in terms of $P$ and $Q$ as:
\ba
R^2_{N}=\theta \sum_{\alpha=0}^{N-1} \left( P_\alpha + Q_\alpha \right) = \theta \left( \sum_{\alpha=0}^{N-1} 2 P_\alpha + N \right)= \theta \left( \sum_{\alpha=0}^{N-1} 2 Q_\alpha - N \right) \ .
\label{RandPQ}
\ea
Finally if we make a quasi-hole of charge $q$ in the origin of the droplet, due to the boundary conditions (\ref{bound}), we obtain $(\sum_{\alpha=0}^{N-1} P_\alpha)^{(q)} = (\sum_{\alpha=0}^{N-1} P_\alpha)^{(0)}  + \alpha_1$ , $(\sum_{\alpha=0}^{N-1} Q_\alpha)^{(q)} = (\sum_{\alpha=0}^{N-1} Q_\alpha)^{(0)} + \alpha_2$ and from (\ref{rel2}) and (\ref{RandPQ}) we arrive to:
\ba
\alpha_1=\alpha_2= q \ .
\label{windcharg}
\ea
It is clear now why we introduce the factor $2\theta$ in the change of variables (\ref{newvar}). Without this factor relation (\ref{windcharg}) should be $\alpha_1=\alpha_2=2\theta q$. But we prefer to express the winding number in unity of charge to connect with the physics of the chiral boson, as will be clear later.

Now we come back to the Lagrangian (\ref{GaussBound}). In appendix A.1, we show that solving the constraint (\ref{GaussBound1}) and integrating out the $Q$ and $T$ fields in (\ref{GaussBound}), the following Lagrangian is obtained:
\ba
L_b= 2\theta \sum_{\alpha=0}^{N-1} P_\alpha \dot{R}_\alpha + 2\theta \sum_{\alpha=0}^{N-1} \sum_{\mu \neq \alpha} P_\mu \dot{U}_\alpha - \theta^2 \omega \left( \sum_{\alpha=0}^{N-1} P_\alpha \right)^2 \ .
\label{lagran2}
\ea
In (\ref{lagran2}) an auxiliary field $U$, satisfying $U_{N-1}=U_{0} + \beta_3$, has been introduced as shown in appendix A.1. We have introduced this trivial term because in this way the Lagrangian (\ref{lagran2}) has an $U(1)$ gauge invariance
given by:
\ba
R_\alpha (t) \rightarrow R_\alpha (t) +
\lambda_\alpha (t) \ , \ U_\alpha (t) \rightarrow
U_\alpha (t) + \lambda_\alpha (t) \ , \ \alpha=0,..,N-1 \ ,
\label{gauge}
\ea
where $\lambda_\alpha$ is an arbitrary function of time. It is easy
to see that the transformations (\ref{gauge}) produce the following
change in the Lagrangian:
\ba
L_b^\lambda = L_b + \sum_{\alpha=0}^{N-1}
\dot{\lambda}_\alpha \sum_{\beta=0}^{N-1} P_\beta \ ,
\label{changegauge}
\ea
and because $\sum_{\beta=0}^{N-1} P_\beta$ commutes with the Hamiltonian,
the last term in (\ref{changegauge}) is a total time derivative and therefore
the Lagrangian (\ref{lagran2}) is gauge invariant.

Finally the path integral can be written as:
\ba
Z &=& \prod_{\alpha=0}^{N-1} \int DP_\alpha DU_\alpha DR_\alpha  \ e^{-\int dt L_b} \ \delta (P_{N-1}-P_0+\alpha_1) \delta (U_{N-1}-U_0+\beta_3) \nl && \qquad \delta (R_{N-1}-R_0+\beta_1),
\label{pathinte}
\ea
with $L_b$ given by (\ref{lagran2}) and satisfying the gauge invariance (\ref{gauge}).

Later we will fix the gauge but first it is better to perform another change of variables.
As said in the last of section (2.1) the low energy excitations of the CS matrix theory, described on the boundary of the droplet by the effective theory (\ref{pathinte}), are in one to one correspondence with the excitations of the chiral boson theory. Therefore we want to connect the boundary theory (\ref{pathinte}) with the physics of the chiral boson. To do this it is better follows the approach of Floreanini and Jackiw in Ref.\cite{jack} and write the chiral boson Lagrangian in terms of the boson field $\chi(x)$ as:
\ba
L=\frac{1}{4}\int dx dy \chi (x) \epsilon(x-y) \dot{\chi}(y) - \frac{1}{2} \int dx \chi(x)^2,
\label{boson-FJ}
\ea
where $\epsilon(x)$ is the unit step function. If we define the non-local field $\Phi(x)=\int dy \epsilon(x-y) \chi(y)$ we can write (\ref{boson-FJ}) as the typical chiral boson Lagrangian:
\ba
L=\frac{1}{2} \int dx \dot{\Phi}(x) \Phi'(x) -\frac{1}{2} \int dx \Phi'^2(x) \ .
\label{boson-FJ1}
\ea
Based on the previous argument we perform another change of variables in the path
integral (\ref{pathinte}) given by:
\ba
P_\alpha \rightarrow \widetilde{P}_\alpha= \sum_{\mu
\neq \alpha}^{N-1} P_{\mu}=P^>_\alpha + P^<_\alpha \ , \ \alpha=0,..,N-1 \ ,
\label{change}
\ea
where
\ba
P_\alpha^>=\sum_{\mu>\alpha}^{N-1} P_\mu \
\quad \mbox{and}  \quad P_\alpha^<=\sum_{\mu=0}^{\mu<\alpha} P_\mu \ , \quad
\alpha=0,...,N-1 \ .
\label{Pes}
\ea
Observe that in the continuum limit, that will be taken later, the variables (\ref{Pes}) become analogous to the field $\Phi(x)$ in (\ref{boson-FJ1}), i.e. $P^>_\alpha \rightarrow P^>(x)=\int dy \epsilon(y-x) P(y)$ and $P^<_\alpha \rightarrow P^<(x)=\int dy \epsilon(x-y) P(y)$.

Now from (\ref{change}) and (\ref{Pes}) we obtain the following constraints:
\ba
P_i &=& P^<_{i+1} - P^<_{i} = P^>_{i-1}-P^>_i \ , \ i=1,..,N-2 \ , \quad P_0=P_1^< \ , \quad P_{N-1}=P^>_{N-2} \nl
P^>_{0} &=& P^<_{N-1} + \alpha_1 \ ,
\label{const20}
\ea
where the last equality is satisfied due to the boundary condition on the $P$ field, $P_{N-1}=P_0 + \alpha_1$.
Relations (\ref{const20}) allow us to express the effective theory (\ref{pathinte}) in the variables $P^>$ and $P^<$. In terms of the new coordinates the Lagrangian (\ref{lagran2}) is given by:
\ba
L_b&=& 2 \theta \sum_{\alpha=1}^{N-2} ( P^<_{\alpha +1} - P^<_\alpha ) \dot{R}_\alpha + 2\theta \sum_{\alpha=1}^{N-2} (P^>_\alpha + P^<_\alpha) \dot{U}_\alpha - \theta^2 \omega \sum_{\alpha=1}^{N-2} ( P^<_{\alpha + 1} - P^<_{\alpha} )^2  \nl &-& \theta^2 \omega \sum_{\alpha=1}^{N-2} ( P^<_{\alpha + 1} - P^<_{\alpha} ) ( P^<_\alpha + P^>_\alpha ) \ .
\label{lagran1}
\ea
Note that in (\ref{lagran1}) we have excluded the points $\alpha=0$ and $\alpha=N-1$. We do this because we want to express the Lagrangian, $L_b$, in terms of the discrete derivative of $P^<$. In the last of the section we will take the large $N$ limit in which the point $\alpha=1$ goes to $\alpha=0$ and the point $\alpha=N-2$ goes to $\alpha=N-1$ and therefore we will recover the sum over all points of the lattice.

At this point it is convenient to fix the gauge and eliminate the coordinates $R_\alpha$ in the Lagrangian (\ref{lagran1}), as follows:
\ba
R_\alpha^\lambda &=& R_\alpha + \lambda_\alpha= -  P^<_\alpha
\Rightarrow \lambda_\alpha=-  P^<_\alpha - R_\alpha \nl
U_\alpha^\lambda &=& U_\alpha +
\lambda_\alpha = U_\alpha -  P^<_\alpha - R_\alpha
\ , \ \alpha=1,..,N-2 \ .
\label{gaugefix}
\ea
If we now integrate out the coordinates $U^\lambda$'s, we obtain a new constraint on $P^>$ and $P^<$ given by:
\ba
\frac{d \left( P^>_\alpha + P^<_\alpha \right)}{dt}=0 \ , \ \alpha=0,...,N-2 .
\label{const5}
\ea
Finally to obtain the integration measure in terms of $P^>_\alpha$ and $P^<_\alpha$ ($\alpha=0,..,N-1$) note that these variables, for each value of $\alpha$, are independent by construction (\ref{Pes})
and because the integration is in all $\widetilde{P}_\alpha = P_\alpha^> +
P_\alpha^<$ we obtain:
\ba
DP_\alpha=C \ D
\widetilde{P}_\alpha = C \ D P_\alpha^> D P_\alpha^< \ ; \ \alpha=0,..,N-1  \ ,
\ea
where $C$ is a constant.

In conclusion, in the gauge fixing (\ref{gaugefix}) the effective boundary theory expressed in terms of $P^>$ and $P^<$ is given by:
\ba
Z&=&\prod_{\alpha=0}^{N-1} \int DP^>_\alpha DP^<_\alpha \ e^{-\int dt L_b} \ \prod_{\beta=1}^{N-2} \delta\left( \frac{d(P^>_\beta + P^<_\beta)}{dt} \right) \ \prod_{\gamma=1}^{N-2} \delta \left( P^<_{\gamma+1} - P^<_{\gamma}-P^>_{\gamma-1}+P^>_\gamma \right) \nl && \delta(P^<_0 - P^<_{N-1} - \alpha_1) \ ,
\label{pathint}
\ea
with
\ba
L_b&=&- 2 \theta \sum_{\alpha=1}^{N-2} (P^<_{\alpha+1} - P^<_\alpha) \dot{P}^<_\alpha - \theta^2 \omega \sum_{\alpha=1}^{N-2} (P^<_{\alpha+1} - P^<_\alpha)^2 - \theta^2 \omega \sum_{\alpha=1}^{N-2} (P^<_{\alpha+1} - P^<_\alpha) (P^>_\alpha + P^<_\alpha) \ , \nl
\label{efflagran}
\ea
where the delta functions in (\ref{pathint}) correspond to the constraints (\ref{const20}) and (\ref{const5}) and we have defined the zero component of the field $P^<$ as $P^<_0=P^>_0$.

Finally we consider the $N \rightarrow \infty$ limit of (\ref{pathint}) scaling the
$\alpha$ index as $\alpha \rightarrow a \alpha = x$, with $a \simeq
\frac{1}{N}$ the typical spacing of the lattice. Therefore, in the $a
\rightarrow 0$ limit, $a \alpha = x$ becomes a continuous variable
taking values along the interval $[0,1]$. In appendix A.2. we show that in this limit the field $P^>$ can be integrated out and the theory (\ref{pathint}) becomes the chiral boson theory on the unitary circle given by the Lagrangian:
\ba
L_b = - k\int dx  \frac{\partial
P^<(x,t)}{\partial x} \dot{P^<}(x,t)  - \omega' \int dx  \left(
\frac{\partial P^<(x,t)}{\partial x} \right)^2 \ ,
\label{FinalLagrang1}
\ea
and the boundary condition on the $P^<(x,t)$ field:
\ba
P^<(0,t)=P^<(1,t) + \alpha_1 \ ,
\label{zeromode1}
\ea
where $\alpha_1$ (\ref{windcharg}) is equal to the charge of the quasi-hole and $k=B\theta$ is the inverse of the filling fraction of the Laughlin fluid as shown in section (2.1).

In conclusion, starting from the CS theory describing the Laughlin fluids, we were able to identify the bulk and boundary d.o.f. and therefore we could froze the bulk dynamics by means of the mean field (\ref{meanfield2}) obtaining an effective theory for the edge d.o.f. In this section we have obtained that this effective theory is the chiral boson theory on the circle (\ref{FinalLagrang1}). Moreover we have shown that the zero mode $\alpha_1$ (\ref{windcharg}, \ref{zeromode1}), in the boundary theory, is equal to the charge of a quasi-hole (\ref{quasip-inorig}) in the bulk of the fluid and also that the constant $k=B\theta$ in (\ref{FinalLagrang1}) is the inverse of the filling fraction of the ground state (Laughlin wave function) of the CS theory (see section 2).

To conclude the section we want to remark that it will be interesting to connect our analysis with that of Ref.\cite{cabra} in which the authors analyze the chiral boson theory with an additional self-interacting term. In our approach it is possible to introduce interacting terms in the Lagrangian (\ref{FinalLagrang1}) considering more general potentials of that introduced in (\ref{LCS}). This will be study in a future paper.

\section{Conclusions}
In this paper we have studied the boundary physics of the regularized non-commutative CS theory. As we have shown, in section 2.1, the quantum ground state of this theory corresponds to the Laughlin wave function for the FQHE \cite{cr}-\cite{frad} with filling fraction $\nu=\frac{1}{k}$ with $k$ an integer. We also show, in the last of section 2.1, that the low energy excitations (edge excitations) of the CS theory are in one to one correspondence with those of the chiral boson theory that are a very good description of the boundary physics of the FQHE \cite{wen}\cite{winf}. In section 3.1, we have considered a perturbation of the semiclassical droplet solution that made possible to identify the relevant d.o.f. on the edge. Based on this analysis we introduced a physical mean field approximation that froze the bulk d.o.f. and their interactions with the boundary, to the semiclassical values, obtaining an effective theory for the edge of the fluid. This effective theory, as shown in section 3.2, is the chiral boson theory on the circle (\ref{FinalLagrang1}).
We have also obtained that, in the edge theory (\ref{FinalLagrang1}), the zero mode $\alpha_1$ (\ref{windcharg}, \ref{zeromode1}) corresponds to the charge of a quasi-hole in the bulk theory (\ref{CS-poly-holom}) and the constant $k$ is the inverse of the filling fraction of the ground state of the CS matrix theory given, as said before, by the Laughlin wave function (see section 2.1).

In a future paper we plan to do a similar analysis in the Maxwell CS matrix theory. As shown in \cite{we}\cite{we2} it is a good framework to study the Hierarchical Jain states. For this reason we expect a connection between the boundary excitations of the Maxwell CS matrix theory and the $W_\infty$ conformal field theories for the FQHE \cite{winf}\cite{jainedge}. Also it should be interesting to consider in our analysis a mean field approximation that takes into account interactions between the bulk and the boundary.

\newpage

{\large \bf Acknowledgments}

I would like to thank A. Cappelli for many discussions and suggestions on this work and also to D. Seminara, G.Viola, V.Guarrera, G.Barontini, J.Bechi and V.Cardinali for helpful conversations. Also I would like to thank the Referee of the JHEP journal for many valuable comments and suggestions.
At last I thank the Galileo Galilei Institute for Theoretical Physics for the hospitality, the ESF Science Programme INSTANS 2005-2010 and the INFN for partial support during the completion of this work.

\appendix

\section{Appendix}

\subsection{}
In this appendix after to solve the constraint (\ref{GaussBound1}):
\ba
G_{NN} =  \sum_{\alpha=0}^{N-1}
\left( P_\alpha - Q_\alpha  \right) +  N  = 0 \ ,
\label{GaussBound1a}
\ea
we are able to integrate out the $Q$ and $T$ fields in the Lagrangian (\ref{GaussBound}) given by:
\ba
L_b = 2\theta \sum_{\alpha=0}^{N-1} (P_\alpha \dot{R}_\alpha + Q_\alpha \dot{T}_\alpha) - J_0 \omega\theta \sum_{\alpha=0}^{N-1} (P_\alpha + Q_\alpha) - \theta^2 \omega \left( \sum_{\alpha=0}^{N-1} (P_\alpha + Q_\alpha) \right)^2 \ .
\label{GaussBounda}
\ea
First we eliminate the term $J_0 \omega \theta \sum_{\alpha=0}^{N-1} (P_\alpha + Q_\alpha)$ in (\ref{GaussBounda}), by means of a rigid translation of the $P$ (or $Q$) field, i.e. $P_\alpha \rightarrow P_\alpha + a$, $\alpha=0,..,N-1$ with $a$ a constant. At this point we solve the constraint (\ref{GaussBound1a}) in term of one of the coordinates, for instance $Q_0$, obtaining:
\ba
Q_0=\sum_{\beta=0}^{N-1} P_\beta - \sum_{\mu>0} Q_\mu + N \ .
\label{SolConst}
\ea
After integration of this variable in the path integral we arrive to the following Lagrangian:
\ba
L_b= 2\theta \sum_{\alpha=0}^{N-1} P_\alpha \dot{R}_\alpha + 2\theta \left( \sum_{\beta=0}^{N-1} P_\beta - \sum_{\mu>0} Q_\mu + N \right) \dot{T}_0 + \sum_{\mu>0}Q_\mu \dot{T}_\mu - \theta^2 \omega \left( \sum_{\alpha=0}^{N-1} P_\alpha + N \right)^2. \nl
\label{lagran}
\ea
As we did before we can do a rigid translation of the $P$ field and write the Hamiltonian as:
\ba
H=\theta^2\omega \left( \sum_{\alpha=0}^{N-1} P_\alpha \right)^2 \ .
\label{hamil}
\ea
Finally if we integrate out the $Q$ and $T$ fields we obtain the following Lagrangian:
\ba
L_b=2\theta \sum_{\alpha=0}^{N-1} P_\alpha \dot{R}_\alpha - \theta^2\omega \left( \sum_{\alpha=0}^{N-1} P_\alpha \right)^2 \ .
\label{lagrang1}
\ea
Now we will rewrite the Lagrangian (\ref{lagrang1}) in another way. Using the fact that $\sum_{\alpha=0}^{N-1} P_\alpha$ is a constant of motion we can add a zero term $2\theta \sum_{\alpha=0}^{N-1} \dot{U}_\alpha \sum_{\beta=0}^{N-1} P_\beta$ to (\ref{lagrang1}), by means of an auxiliary field $U_\alpha$, $\alpha=0,...,N-1$ verifying an arbitrary boundary condition $U_{N-1}=U_0 + \beta_3$, with $\beta_3$ a constant. Therefore (\ref{lagrang1}) becomes:
\ba
L_b= 2\theta \sum_{\alpha=0}^{N-1} P_\alpha \dot{R}_\alpha + 2\theta \sum_{\alpha=0}^{N-1} \sum_{\mu \neq \alpha} P_\mu \dot{U}_\alpha - \theta^2 \omega \left( \sum_{\alpha=0}^{N-1} P_\alpha \right)^2 \ ,
\label{lagran2a}
\ea
where in (\ref{lagran2a}) we have absorbed the field $U$ in a redefinition of the field $R$, i.e. $R_\alpha \rightarrow R_\alpha + U_\alpha$, $\alpha=0,...,N-1$.

\subsection{}
In this appendix we consider the limit $N \rightarrow \infty$ of the theory (\ref{pathint}) given by:
\ba
Z&=&\prod_{\alpha=0}^{N-1} \int DP^>_\alpha DP^<_\alpha \ e^{-\int dt L_b} \ \prod_{\beta=1}^{N-2} \delta\left( \frac{d(P^>_\beta + P^<_\beta)}{dt} \right) \ \prod_{\gamma=1}^{N-2} \delta \left( P^<_{\gamma+1} - P^<_{\gamma}-P^>_{\gamma-1}+P^>_\gamma \right) \nl && \delta(P^<_0 - P^<_{N-1} - \alpha_1) \ ,
\label{pathinta}
\ea
with
\ba
L_b&=&- 2 \theta \sum_{\alpha=1}^{N-2} (P^<_{\alpha+1} - P^<_\alpha) \dot{P}^<_\alpha - \theta^2 \omega \sum_{\alpha=1}^{N-2} (P^<_{\alpha+1} - P^<_\alpha)^2 - \theta^2 \omega \sum_{\alpha=1}^{N-2} (P^<_{\alpha+1} - P^<_\alpha) (P^>_\alpha + P^<_\alpha) \ . \nl
\label{efflagrana}
\ea
In this limit the $\alpha$ index scale as $\alpha \rightarrow a \alpha$, with $a \simeq
\frac{1}{N}$ the typical spacing of the lattice. So, in the $a
\rightarrow 0$ limit, $a \alpha = x$ becomes a continuous variable
taking values along the interval $(0,1)$, and we obtain that:
\ba
&& P^<_\alpha \rightarrow P^<(x,t) \ , \ P^>_\alpha \rightarrow P^>(x,t)
\ , \ \sum_{\alpha=1}^{N-2}
\rightarrow \lim_{a\rightarrow 0} \frac{1}{a} \int_0^1  dx \ , \ \nl &&
P^<_{\alpha + 1} - P^<_\alpha \rightarrow a \frac{\partial
P^<(x,t)}{\partial x} \ , \  P^>_\alpha - P^>_{\alpha-1} \rightarrow a
\frac{\partial P^>(x,t)}{\partial x} \ .
\label{contlimit}
\ea
Therefore using (\ref{contlimit}) the Lagrangian (\ref{efflagrana}) becomes:
\ba
L_b &=& - 2 \theta \int_0^1 dx \frac{\partial P^<(x,t)}{\partial x} \dot{P^<}(x,t)  -  a \theta^2 \omega \int_0^1 dx
\left( \frac{\partial P^<(x,t)}{\partial x} \right)^2 \nl &-& \theta^2 \omega
\int_0^1 dx \frac{\partial P^<(x,t)}{\partial x} \left( P^>(x,t) + P^<(x,t)
\right) ,
\label{quasiL}
\ea
and the constraints in the path integral (\ref{pathinta}) are:
\ba
\frac{\partial P^<(x,t)}{\partial x} &=& -
\frac{\partial P^>(x,t)}{\partial x} \ , \ \frac{d\left(P^>(x,t)+P^<(x,t) \right)}{dt}=0 \Rightarrow P^<(x,t) + P^>(x,t) = c \ ; \nl P^<(0,t) &=& P^<(1,t) + \alpha_1 \ ,
\label{contconstr}
\ea
where $c$ is a constant. Thus in the continuum limit the fields $P^>$ and $P^<$ are related by a constant.

Now using (\ref{contconstr}) into (\ref{quasiL}) and restoring the $B$ dependence we obtain the chiral boson Lagrangian on the unitary circle \cite{wen}\cite{winf}:
\ba
L_b = - k\int dx  \frac{\partial
P^<(x,t)}{\partial x} \dot{P^<}(x,t)  - \omega' \int dx  \left(
\frac{\partial P^<(x,t)}{\partial x} \right)^2 \ ,
\label{FinalLagrang}
\ea
with $\omega'=\frac{B}{2}  a \theta^2 \omega$  and the field $P^>(x,t)$ satisfying the boundary condition (\ref{contconstr}):
\ba
P^<(0,t)=P^<(1,t) + \alpha_1 \ .
\label{zeromode}
\ea

\end{document}